\documentclass[12pt]{article}
\usepackage{amssymb}
\usepackage{bm}
\usepackage{mathrsfs}
\usepackage{amsmath}

\def\1{{\bf 1}}
\def\G{{\bf G}}
\def\H{{\bf H}}
\def\K{{\bf K}}
\def\S{{\bf S}}
\def\Gz{\mathbb{G}}
\def\Lz{\mathbb{L}}
\def\ol{\overline}

\begin{document}

\title{Explicit effective Hamiltonians for general linear
quantum-optical networks}
\author{U Leonhardt\\
School of Physics and Astronomy, University of St Andrews,\\
North Haugh, St Andrews, Fife, KY16 9SS, Scotland\\~\\
A Neumaier\\
Institut f\"ur Mathematik, Universit\"at Wien,\\
Strudlhofgasse 4, A-1090 Wien, Austria}
\maketitle
\begin{abstract}
Linear optical networks are devices that turn classical
incident modes by a linear transformation into outgoing ones.
In general, the quantum version of such transformations may mix
annihilation and creation operators. We derive a simple formula
for the effective Hamiltonian of a general linear quantum network,
if such a Hamiltonian exists. 
Otherwise we show how the scattering matrix of the network
is decomposed into a product of three matrices
that can be generated by Hamiltonians.

\end{abstract}

\newpage

Simple optical instruments \cite{LeoReview} such as beam splitters
or parametric amplifiers are characterized by linear input-output
relations.
The beam splitter transforms the annihilation operator of the incident
light modes
according to the classical laws of optical interference, i.e.,
by a linear transformation. The parametric amplifier acts like a
phase-conjugating mirror, combining the annihilation operator of one
incident mode with the creation operator of the other.
Complex optical networks can be constructed from beam-splitters,
mirrors and active elements such as parametric amplifiers
\cite{Reck,Mattle,TormaJex1}.

Networks are essential to the optical communication of 
information and for novel ways of quantum-information 
processing \cite{CAW,KLM}. 
Furthermore, linear networks may possess interesting
quantum-statistical properties when they are large, for instance
as representations of the Ising model \cite{Torma,TormaJex2} and
as examples of quantum localization \cite{TJS}. Here we derive a
simple formula for the Hamiltonian of an arbitrarily large
quantum-optical network, if such a Hamiltonian exists.
Otherwise we show how the scattering matrix of the network
is decomposed into a product of three matrices
that can be generated by Hamiltonians.
Our results allows to predict how the
quantum state of the incident light modes is processed. Our
theory contains as special cases the previously studied
Hamiltonians of symmetric passive networks \cite{TormaJex1}, the
theory of the beam splitter \cite{Campos,LeoSU2,Leonhardt} and of
the parametric amplifier \cite{YMK,LeoSU11}.

Consider $n$ incident modes with the annihilation operators
$\widehat{a}_k$ and the creation operators $\widehat{a}_k^\dagger$,
the index $k$ running from $1$ to $n$. Suppose that the optical
network produces $n'$ outgoing modes with the corresponding
operators $\widehat{a}_k'$ and $\widehat{a}_k'^\dagger$.
(As we shall see in (\ref{eq:nn}) below, $n'$ must equal $n$.)

Canonical quantization of the linear input-output behavior of a
classical linear network leads to the requirement that the mode
operators are related to each other by the linear transformation
\begin{equation}
\label{eq:linear}
\left(
    \begin{array}{c}
     \widehat{a}_k'  \\
     \widehat{a}_k'^\dagger
    \end{array}
\right)
=
\S
\left(
    \begin{array}{c}
     \widehat{a}_k  \\
     \widehat{a}_k^\dagger
    \end{array}
\right)\,.
\end{equation}
(In this formula, $\widehat{a}_k$ is representative for the vector
formed by all annihilation operators, and similar for the other
operators indexed by $k$.)
$\S $ denotes the classical scattering matrix of the network;
it has $2n'$ rows and $2n$ columns and must be of the form
\begin{equation}\label{eq:Sform}
\S = \begin{pmatrix} A & B \\ \ol B & \ol A\end{pmatrix},
\end{equation}
where $\ol A$ denotes the (untransposed) complex conjugate of $A$.
Indeed, (\ref{eq:Sform}) is necessary for the consistency of
(\ref{eq:linear}), as can be seen by writing in components and
forming the adjoint. If $B=0$, the network is passive and can be
constructed from beam splitters, phase shifters and mirrors \cite{Reck}.
Otherwise, the transformation (\ref{eq:linear}) mixes
annihilation and creation operators, and we speak of an active
network.

In the Schr\"odinger picture, incident light in the state $|\psi\rangle$
is transformed into the state $|\psi'\rangle=\widehat S|\psi\rangle$ of
the outgoing light, with a unitary scattering operator $\widehat{S}$.
In the Heisenberg picture, this corresponds to the transformation
$\widehat O' = \widehat{S}^\dagger \widehat O \widehat{S}$
for arbitrary operators $\widehat O$. In particular, to
match the mode transformation (\ref{eq:linear}), we need to have
\begin{equation}\label{eq:bop}
\S
\left(
    \begin{array}{c}
     \widehat{a}_k  \\
     \widehat{a}_k^\dagger
    \end{array}
\right)
=
\widehat{S}^\dagger
\left(
    \begin{array}{c}
     \widehat{a}_k  \\
     \widehat{a}_k^\dagger
    \end{array}
\right)
\widehat{S}.
\end{equation}
Here it is understood that $\widehat S$ acts on each creation and
annihilation operator separately. In the following we show how to
construct from the classical scattering matrix $\S $ a unitary
scattering operator $\widehat{S}$ such that (\ref{eq:bop}) holds;
this operator then completely specifies the desired quantum
behavior of the network.

The light of both the incident and the
outgoing modes consists of bosons subject to the commutation rules
\begin{eqnarray}
\label{eq:comm}
[\widehat{a}_k, \widehat{a}_{k'}^\dagger] =  \delta_{kk'}\,,
&& \quad
[\widehat{a}_k, \widehat{a}_{k'}] =  0\,,
\\ \nonumber
[\widehat{a}_k', \widehat{a}_{k'}'^\dagger] =  \delta_{kk'}\,,
&& \quad
[\widehat{a}_k', \widehat{a}_{k'}'] =  0\,.
\end{eqnarray}
We write the commutation relations (\ref{eq:comm}) in matrix form,
\begin{eqnarray}
\Big[\left(
    \begin{array}{c}
     \widehat{a}_k  \\
     \widehat{a}_k^\dagger
    \end{array}
\right),
(\widehat{a}_{k'}^\dagger, \widehat{a}_{k'})
\Big]
&=&
\left(
    \begin{array}{cc}
      {[\widehat{a}_k,  \widehat{a}_{k'}^\dagger]} &
      {[\widehat{a}_k,  \widehat{a}_{k'}]} \\
      {[\widehat{a}_k^\dagger,  \widehat{a}_{k'}^\dagger]}  &
      {[\widehat{a}_k^\dagger,  \widehat{a}_{k'}]}
    \end{array}
\right)
= \G\,,
\nonumber\\
\Big[\left(
    \begin{array}{c}
     \widehat{a}_k'  \\
     \widehat{a}_k'^\dagger
    \end{array}
\right),
(\widehat{a}_{k'}'^\dagger, \widehat{a}_{k'}')
\Big]
&=&
\left(
    \begin{array}{cc}
      {[\widehat{a}_k',  \widehat{a}_{k'}'^\dagger]} &
      {[\widehat{a}_k',  \widehat{a}_{k'}']} \\
      {[\widehat{a}_k'^\dagger,  \widehat{a}_{k'}'^\dagger]}  &
      {[\widehat{a}_k'^\dagger,  \widehat{a}_{k'}']}
    \end{array}
\right)
= \G '
\label{eq:c}
\end{eqnarray}
with
\begin{equation}\label{eq:G}
\G  =
\left(
    \begin{array}{cc}
      \1  & 0 \\
       0 & -\1
    \end{array}
\right)
\,,\quad
\G ' =
\left(
    \begin{array}{cc}
      \1 ' & 0 \\
       0 & -\1 '
    \end{array}
\right)\,,
\end{equation}
where $\1 $ and $\1 '$ denote the unit matrix in $n$ and $n'$
dimensions, respectively. Inserting the linear mode transformation
(\ref{eq:linear}) into the commutation relations and using
(\ref{eq:Sform}), we find that the commutation relations are
equivalent to $\G'  = \S \,\G \,\S ^\dagger$.
This relation implies that $\G\S ^\dagger \G'$
is the inverse of $\S $. Since $\S $ is invertible,
the number of incident modes must equal the number of outgoing modes,
\begin{equation}\label{eq:nn}
n = n',
\end{equation}
a result well-known for passive networks made of beam splitters
and mirrors \cite{Reck,Mattle,TormaJex1}.
This feature therefore remains true for general quantum-optical
networks.
In particular, even if just one light beam is transformed into
two or more beams a matching number of incident vacuum modes are
involved with their vacuum noise affecting the outgoing light.

Since (\ref{eq:nn}) implies $\G'=\G$, we find
\begin{equation}
\label{eq:qu}
 \G= \S \,\G \,\S ^\dagger \,
\end{equation}
as another consistency condition for classical scattering matrices.
The matrices $\S$ satisfying (\ref{eq:qu}) are called quasi-unitary
\cite{Cornwell} and form a Lie group $\Gz_0$
whose infinitesimal generators are the elements of the quasi-unitary
Lie algebra $\Lz_0$ consisting of all matrices $\K$ with
\begin{equation}\label{eq:lie}
\K\G+\G \K^\dagger =0.
\end{equation}
Because of (\ref{eq:G}), (\ref{eq:lie}) implies that the elements
of $\Lz_0$ are precisely those of the form
\begin{equation}\label{eq:lieform}
\K = \begin{pmatrix} A & D \\ D^\dagger & B\end{pmatrix}
\end{equation}
with antihermitian $A, B$ and arbitrary $D$.
Because of (\ref{eq:Sform}), an $\S$ suitable as a classical
scattering matrix in fact belongs to a subgroup
$\Gz$ of $\Gz_0$; the corresponding Lie subalgebra $\Lz$ consists
of all $\K$ of the form
\begin{equation}\label{eq:lieform2}
\K = \begin{pmatrix} A & D \\ \ol D & \ol A\end{pmatrix}
\end{equation}
with antihermitian $A$ and complex symmetric $D$.

By standard results for Lie groups, every element $\S\in\Gz$
can be written (in many ways) as a product
\begin{equation}\label{eq:product}
\S=e^{\K_1}\cdots e^{\K_m}
\end{equation}
of finitely many exponentials of infinitesimal generators
$\K_1 , \dots, \K_m \in \Lz$.

A special case in which a single exponential often suffices is when
$\S\in\Gz$ can be diagonalized, i.e., $\S=X\Lambda X^{-1}$
with a diagonal matrix $\Lambda$, and has no eigenvalues that are
real and negative. In this case $\S=e^\K$ with $\K=X(\ln\Lambda)X^{-1}$,
where, for the logarithm $\ln \Lambda$, the principal value is taken
in each diagonal element. Often, and if $\K$ is sufficiently small,
always, $\K\in\Lz$; then (\ref{eq:lie}) holds.

For general $\S\in\Gz$, a decomposition (\ref{eq:product}) of the form
\begin{eqnarray}
\S&=&\exp\begin{pmatrix} A_{1} & 0 \\ 0 & \ol A_{1}\end{pmatrix}
\exp\begin{pmatrix} 0& D \\ D  & 0\end{pmatrix}
\exp\begin{pmatrix} A_{3} & 0 \\ 0 & \ol A_{3}\end{pmatrix}
\nonumber\\
&=& \begin{pmatrix} \exp {A_{1}} & 0 \\ 0 &
\ol{\exp{A_{1}}}\end{pmatrix}
\begin{pmatrix} \cosh D & \sinh D \\ \sinh D  & \cosh D\end{pmatrix}
\begin{pmatrix} \exp{A_{3}} & 0 \\ 0 & \ol{\exp{A_{3}}}\end{pmatrix},
\label{eq:Sfact}
\end{eqnarray}
with antihermitian $A_{j}$ and real diagonal $D$ can be found
constructively (see Appendix A); the resulting $\K_j$ belong to
$\Lz$ since they have the form (\ref{eq:lieform2}).
The factorization can be given a natural physical interpretation,
similar to the one of the known factorization of two-mode parametric
amplifiers \cite{LeoReview,YMK,LeoSU11}:
Any linear network can be thought of acting in three steps.
First the incident modes are mixed by a passive network,
then they undergo parametric amplification with real squeezing
parameters, and finally they are subject to another passive mode
transformation. In the special case when the entire
network is passive the squeezing parameters are zero.

Given the structure of general group elements $\S$, we may
concentrate in the construction of the unitary scattering operator
$\widehat{S}$ satisfying (\ref{eq:bop}) on the case
of a single exponential $\S=e^\K$ with $\K\in\Lz$.
We observe that the operator
\begin{equation}\label{eq:ham}
\widehat H_\K =
\frac{1}{2}\,(\widehat{a}_k^\dagger, \widehat{a}_k)\, \H  \left(
    \begin{array}{c}
     \widehat{a}_k  \\
     \widehat{a}_k^\dagger
    \end{array}
\right) ~~~\mbox{ with } \H = -i\G \K
\end{equation}
is Hermitian. Indeed, $\H$ is Hermitian since
\begin{equation}
\H^\dagger = i\K^\dagger \G^\dagger =i\K^\dagger \G=-i\G \K=\H
\end{equation}
by (\ref{eq:lie}), and the hermiticity of $\widehat H_\K$ follows.
The operator
$\widehat H_\K $ plays the role of an effective Hamiltonian in a
fictitious time evolution for the quantum-optical network
\cite{TormaJex1},
generating the linear mode transformation (\ref{eq:linear}) in the
Heisenberg picture.
(In simple cases,
the fictitious time $\tau$
corresponds to the fractional depth of the optical device.)
To show this we put
\begin{equation}\label{eq:Stau}
\widehat S(\tau)=\exp(-i\tau\widehat H_\K )
\end{equation}
and verify that the operator vectors
\begin{equation}\label{eq:AB}
\widehat A(\tau) := e^{\tau \K}
\left(
    \begin{array}{c}
     \widehat{a}_k  \\
     \widehat{a}_k^\dagger
    \end{array}
\right),~~~
\widetilde A(\tau):=\widehat S(\tau)^\dagger
\left(
    \begin{array}{c}
     \widehat{a}_k  \\
     \widehat{a}_k^\dagger
    \end{array}
\right)
\widehat S(\tau).
\end{equation}
satisfy the same differential equation
\begin{equation}\label{eq:diffeq}
\frac{d}{d\tau}\widehat A(\tau)= \K\widehat A(\tau).
\end{equation}
Indeed, (\ref{eq:diffeq}) holds trivially for $\widehat A(\tau)$.
To show that it also holds for $\widetilde A(\tau)$, we use
the commutation relations in matrix form (\ref{eq:c}) and get,
using (\ref{eq:Stau}),
\begin{eqnarray}
\frac{d}{d\tau}\widetilde A(\tau)
&=& i\exp(i\tau\widehat H_\K )\,
\Big[\widehat H_\K , \left(
    \begin{array}{c}
     \widehat{a}_k  \\
     \widehat{a}_k^\dagger
    \end{array}
\right) \Big] \exp(-i\tau\widehat H_\K )
\nonumber\\
&=&  i\exp(i\tau\widehat H_\K )\,
 \G \,\H  \left(
    \begin{array}{c}
     \widehat{a}_k  \\
     \widehat{a}_k^\dagger
    \end{array}
\right)
\exp(-i\tau\widehat H_\K )
\nonumber\\
&=& i\G \,\H \,\widehat S(\tau)^\dagger  \left(
    \begin{array}{c}
     \widehat{a}_k  \\
     \widehat{a}_k^\dagger
    \end{array}
\right) \widehat S(\tau)
\nonumber\\
&=&i\G\H\widetilde A(\tau)
=\K\widetilde A(\tau)
\,,
\end{eqnarray}
so that the differential equation (\ref{eq:diffeq}) also holds with
$\widetilde A(\tau)$ in place of $\widehat A(\tau)$.
Since $\widehat A(\tau)$ and $\widetilde A(\tau)$ trivially agree at
$\tau=0$, they agree for all $\tau$. In particular, we conclude that
$\widehat A(1)=\widetilde A(1)$. In view of (\ref{eq:Stau}) and
(\ref{eq:AB}), this implies that the unitary operator
\begin{equation}\label{eq:SK}
\widehat S_\K = \exp(-i\widehat H_\K )
\end{equation}
satisfies the required equation (\ref{eq:bop}) for $\S=e^\K$.
Therefore, $\widehat S_\K$ is the desired scattering operator
corresponding to the classical scattering matrix $\S=e^\K$.

In the more general case
where the classical scattering matrix $\S$ is given by a product
(\ref{eq:product}) of exponentials, one sees by direct substitution
that (\ref{eq:bop}) is satisfied by the scattering operator
\begin{equation}
\widehat S = \widehat S_{\K_m}\cdots\widehat S_{\K_1}.
\end{equation}

Our general formula contains as special cases the known effective
Hamiltonians for beam splitters and parametric amplifiers
\cite[Section 3.3]{LeoReview} that are characterized by the
scattering matrices
\begin{equation}
\S_{split}(\phi)
= \begin{pmatrix} \cos\phi & -\sin\phi & 0 & 0 \\ \sin\phi &
\cos\phi & 0 & 0\\ 0 & 0 & \cos\phi & -\sin\phi\\ 0 & 0 &
\sin\phi & \cos\phi
\end{pmatrix}
\end{equation}
with real $\phi$ for a beam splitter and
\begin{equation}
\S_{amp}(\zeta)
= \begin{pmatrix} \cosh\zeta & 0 & 0 & \sinh\zeta\\
0& \cosh\zeta & \sinh\zeta & 0\\ 0& \sinh\zeta & \cosh\zeta & 0\\
\sinh\zeta & 0 & 0 & \cosh\zeta
\end{pmatrix}
\end{equation}
with real $\zeta$ for a parametric amplifier. One can easily
diagonalize the matrices $\S_{split}(\phi)$ and $\S_{amp}(\zeta)$ 
and calculate so their logarithms $\K_{split}$ and $\K_{amp}$. 
However, the product
\begin{equation}
\S = \S_{amp}(\zeta) \S_{split}(\phi) 
\quad\mbox{for}\,\, \cosh\zeta\cos\phi = 1
\end{equation}
does not possess a diagonal representation. This case represents
a beam splitter that is exactly compensated by parametric
amplification. For example it describes an eavesdropping attempt
where quantum information is tapped by beam splitting followed by
amplification to restore the light intensity. To find the
Hamiltonian for this device, one has to resort to a Jordan
decomposition of the scattering matrix. The resulting matrix
logarithm is $\K = \S - \1$. Finally, we mention a simple example
where the logarithm of the scattering matrix fails to satisfy 
the Lie condition (\ref{eq:lie}) that is 
essential for the hermiticity of the Hamiltonian (\ref{eq:ham}).
The example is a single-mode squeezer followed 
by a $\pi$ phase shifter, 
characterized by the scattering matrix
\begin{equation}
\S= -\begin{pmatrix} 
\cosh\zeta & \sinh\zeta\\
\sinh\zeta & \cosh\zeta
\end{pmatrix},~~~\zeta \ne 0.
\label{eq:ex}
\end{equation}
Since the trace of $\S$ is $-2\cosh\zeta<-2$, the matrix $\K=\log \S$ 
must fail to satisfy (\ref{eq:lie}) 
for arbitrary choices of the branch of the logarithms.
Indeed, if (\ref{eq:lie}) holds, $\K$ has the form (\ref{eq:lieform2}) 
with purely imaginary $A$, hence $\K$ has two real 
or two purely imaginary eigenvalues $\lambda_{1,2}$. Thus the trace
of $S = e^\K$ is $e^{\lambda_1}+e^{\lambda_2}$, which cannot be a 
real number $<-2$. Consequently, there is no single Hamiltonian
that generates a single-mode squeezer followed 
by a $\pi$ phase shifter,
although both subdevices possess effective Hamiltonians.

To summarize, we have developed an explicit procedure how to
calculate the Hamiltonian of a quantum-optical linear network. If
the matrix logarithm $\K$ of the scattering matrix $\S$ satisfies
the Lie condition (\ref{eq:lie}), the Hamiltonian is given by
(\ref{eq:ham}). One can calculate $\K$ by diagonalizing $\S$ or,
if this is not possible, using the Jordan decomposition of $\S$.
We have shown that the scattering matrix $\S$ of any linear
quantum-optical network can be decomposed into three factors that
can be diagonalized or are already diagonal. Note that the
Hamiltonian (\ref{eq:ham}) is not unique, because the matrix
logarithm is multivalued. Furthermore, since any decomposition
(\ref{eq:product}) leads to a realization of the network, there
are many equivalent ways to design an optical network with a
particular input-output relation. There are also many ways to
assemble it from the basic building blocks
\cite{Reck,Mattle,TormaJex1}, from beam splitters and parametric
amplifiers.

\section*{Acknowledgments}

U.L. thanks John Cornwell for discussions and acknowledges the
financial support of the Leverhulme Trust. Thanks to Mike Mowbray
for comments that helped to improve the readability of the paper.

\section*{Appendix A}

In this appendix we prove that every $\S\in\Gz_0$ can be written in the
form (\ref{eq:product}) with three factors of the form
\begin{equation}\label{eq:svdfact}
\K_2=\begin{pmatrix} 0 & D \\ D & 0 \end{pmatrix},~~~
\K_j=\begin{pmatrix} A_{j1} & 0 \\ 0 & A_{j2}\end{pmatrix}
\mbox{~for~} j=1,3,
\end{equation}
with antihermitian $A_{jk}$ and real nonnegative diagonal $D$.
In case that $\S\in\Gz$, the construction can be modified such that
(\ref{eq:Sfact}) hold with antihermitian
$A_{j}$ and real diagonal $D$.

The proof is constructive and begins by partitioning the
matrix $\S$ into four $n\times n$ submatrices,
\begin{equation}\label{eq:S}
\S=\begin{pmatrix} S_{11} & S_{12} \\ S_{21} & S_{22}\end{pmatrix}.
\end{equation}
The condition (\ref{eq:qu}) which expresses the fact that $\S\in\Gz_0$
implies the equations
\begin{equation}\label{eq:1}
S_{11}S_{11}^\dagger -S_{12}S_{12}^\dagger =\1,
\end{equation}
\begin{equation}\label{eq:2}
S_{11}S_{21}^\dagger =S_{12}S_{22}^\dagger ,
\end{equation}
\begin{equation}\label{eq:3}
S_{21}S_{21}^\dagger -S_{22}S_{22}^\dagger =-\1.
\end{equation}
(\ref{eq:1}) implies that
$\|S_{11}^\dagger x\|^2=x^\dagger S_{11}S_{11}^\dagger x
=x^\dagger S_{12}S_{12}^\dagger x+x^\dagger x
=\|S_{12}^\dagger x\|^2+\|x\|^2>0$ if $x\ne 0$,
and therefore that $S_{11}$ is invertible.
It is always possible to factor $S_{12}$ into a product
$S_{12}=U_1SV_2^\dagger$ (singular value decomposition;
see, e.g., \cite{GolvL}) consisting of unitary matrices $U_1$, $V_2$
and a nonnegative real diagonal matrix $S$. The matrix
$C:=(S^2+1)^{1/2}$ is a real, nonnegative invertible diagonal matrix
commuting with $S$ and satisfying
\begin{equation}\label{eq:CS}
C^2-S^2=1.
\end{equation}
We now form the matrices
\begin{equation}
V_1:=S_{11}^{-1}U_1C,~~~U_2:=S_{22}V_2C^{-1}.
\end{equation}
This immediately gives $S_{11}=U_1CV_1^\dagger $ and
$S_{22}=U_2CV_2^\dagger $;
moreover (\ref{eq:2}) implies that $S_{21}=U_2SV_1^\dagger$.
Inserting this into (\ref{eq:1}) and (\ref{eq:3}) shows that $V_1$
and $U_2$ must be unitary, and insertion into (\ref{eq:S}) gives
\begin{equation}\label{eq:S2}
\S = \begin{pmatrix} U_1CV_1^\dagger  & U_1SV_2^\dagger \\
U_2SV_1^\dagger  & U_2CV_2^\dagger\end{pmatrix}
= \begin{pmatrix} U_1  & 0 \\ 0  & U_2\end{pmatrix}
\begin{pmatrix} C& S \\ S & C\end{pmatrix}
\begin{pmatrix} V_1^\dagger  & 0\\ 0  & V_2^\dagger\end{pmatrix}.
\end{equation}
Using (\ref{eq:qu}), it is easily seen that each factor in this
factorization belongs to $\Gz_0$. Due to their special form,
they can be easily be brought into exponential form.
Indeed, since unitary matrices $U$ are normal and have eigenvalues
of absolute value one only, they have a spectral factorization
$U=Q \exp(i\Phi) Q^\dagger $ with unitary $Q$ and real diagonal
$\Phi$, so that $U=e^K$ with antihermitian $K=iQ\Phi Q^\dagger $.
Therefore we can find antihermitian $A_{jk}$ such that
$U_k=e^{A_{1k}}$ and $V_k^\dagger=e^{A_{3k}}$
for $k=1,3$. If we also define the real diagonal matrix
$D:=\log(C+S)$
with componentwise logarithms (nonnegative since $C\ge 1, S\ge 0$)
on the diagonal, it is easily seen that
\begin{equation}
S=\exp\begin{pmatrix} A_{11} & 0 \\ 0 & A_{12}\end{pmatrix}
\exp\begin{pmatrix} 0& D \\ D^\dagger  & 0\end{pmatrix}
\exp\begin{pmatrix} A_{31} & 0 \\ 0 & A_{32}\end{pmatrix},
\end{equation}
with antihermitian $A_{jk}$, as asserted.

Now suppose that $\S\in\Gz$. Comparing (\ref{eq:Sform}) and
(\ref{eq:S2}), we find that we must have
\begin{equation}
\ol U_1 C \ol V_1^\dagger =U_2CV_2^\dagger,~~~
\ol U_1 S \ol V_2^\dagger =U_2SV_1^\dagger.
\end{equation}
If the diagonal entries of $C$ and $S$ are all distinct, the
singular value decomposition is known to be unique up to a
diagonal matrix of phases. Therefore, there are diagonal matrices
$Q_j$ with $Q_j^\dagger Q_j=1$ such that
\begin{equation}
\ol U_1=U_2Q_1,~~~ Q_1\ol V_1^\dagger=V_2^\dagger,~~~
\ol U_1=U_2Q_2,~~~ Q_2\ol V_2^\dagger =V_1^\dagger.
\end{equation}
Clearly, this implies that $Q:=Q_1=Q_2$ is real diagonal with $Q^2=1$;
in particular, $Q$ has diagonal entries $\pm 1$. Therefore
\begin{equation}\label{eq:S20}
\S = \begin{pmatrix} U_1CV_1^\dagger  & U_1SQ \ol V_1^\dagger \\
\ol U_1QSV_1^\dagger  & \ol U_1QCQ \ol V_1^\dagger\end{pmatrix}
= \begin{pmatrix} U_1  & 0 \\ 0  & \ol U_1\end{pmatrix}
\begin{pmatrix} C& QS \\ QS & C\end{pmatrix}
\begin{pmatrix} V_1^\dagger  & 0\\ 0  & \ol V_1^\dagger\end{pmatrix}.
\end{equation}
As before, we can find antihermitian $A_{j}$ such that
$U_1=e^{A_{1}}$ and $V_1^\dagger=e^{A_{3}}$.
If we also define the real diagonal matrix $D:=\log(C+QS)$
with componentwise logarithms on the diagonal, it is easily seen that
(\ref{eq:Sfact}) holds with antihermitian $A_{j}$ and real
diagonal $D$.

Finally, if some diagonal entries of $C$ or $S$ coincide,
we can perturb $\S$ slightly to remove the degeneracy,
and obtain the same decomposition by a limiting argument.

Note added: An equivalent factorization \cite{Braunstein} came to our attention after finishing the paper. 

\bigskip

\end{document}